\documentclass[preprint,12pt,number]{elsarticle}
\usepackage{natbib}
\usepackage{amsmath}   
\usepackage{graphicx}  
\usepackage{amssymb}
\usepackage{lineno}
\usepackage{color,graphics,epsfig}
\journal{Theoretical Population Biology}

\begin{document}

\def\be{\begin{equation}}
\def\ee{\end{equation}}
\def\bea{\begin{eqnarray}}
\def\eea{\end{eqnarray}}
\def\bml{\begin{mathletters}}
\def\eml{\end{mathletters}}
\def\l{\label}
\def\b{\bullet}
\def\eqn#1{(~\ref{eq:#1}~)}
\def\no{\nonumber}
\def\av#1{{\langle  #1 \rangle}}
\newcommand{\tc}{\textcolor}
\begin{frontmatter}
\title{Deterministic evolution of an asexual population under the action of  beneficial and deleterious mutations on additive fitness landscapes}

\author{Kavita Jain and Sona John}

\address{Theoretical Sciences Unit,
Jawaharlal Nehru Centre for Advanced Scientific Research, Jakkur P.O.,
Bangalore 560064, India} 

\begin{abstract}
We study a continuous time model for the frequency distribution of an infinitely large asexual population in which both beneficial and deleterious mutations occur and fitness is additive. When beneficial mutations are ignored, the exact solution for the frequency distribution is known to be a Poisson distribution. Here we include beneficial mutations and obtain exact expressions for the frequency distribution at all times using an eigenfunction expansion method. We find that the stationary distribution is non-Poissonian and related to the Bessel function of the first kind. We also provide suitable approximations for the stationary distribution and the time to relax to the steady state. Our exact results, especially at mutation-selection equilibrium, can be useful in developing semi-deterministic approaches to understand stochastic evolution. 
\end{abstract}

\end{frontmatter}

\section{Introduction}
\label{Intro}

Real populations are finite and evolve stochastically. However theoretical investigations of infinitely large populations subject to various evolutionary forces have been useful in developing diffusion theory, theory of branching process  and semi-deterministic approaches  to describe the evolution of finite populations \cite{Jain:2011c}. Therefore, a detailed analytical study of deterministic models constitutes an important step towards an understanding of  more complex and realistic situations. In the simplest scenario, one may consider an asexual population of infinite size under mutation and selection. 
When the mutation rates depend linearly on the number of deleterious mutations in a sequence and the fitness of the genetic sequence is additive, a complete solution of the genotypic frequency distribution is known exactly \cite{Woodcock:1996,John:2015} and has been utilised, for example, in modeling codon usage bias \cite{Li:1987,Bulmer:1991,Mcvean:1999}.

Another model in which the fitnesses are non-epistatic but the rate of deleterious and beneficial mutations is independent of the fitness of the sequence has recently appeared in various contexts such as adaptive evolution  \cite{Tsimring:1996,Kessler:1997,Desai:2007a,Rouzine:2008,Park:2010}, evolution of sex \cite{Goyal:2012} and evolution of mutation rates \cite{James:2016}. For this mutation scheme, when beneficial mutations are ignored, the exact solution in the stationary state and for the dynamics of the population fraction is known; in particular, at mutation-selection balance, the frequency is Poisson-distributed with a mean given by the ratio of the deleterious mutation rate to selection coefficient \cite{Kimura:1966,Haigh:1978,Maia:2003,Etheridge:2009}.  
When beneficial mutations are also allowed, the short time dynamics of the frequency distribution have been quite well studied \cite{Park:2010} and some approximate results at mutation-selection equilibrium were obtained recently \cite{James:2016}. Here we study this model in detail and find an exact expression for the frequency distribution at all times.

In the following section, we define the model, discuss some limiting cases and explain the relation of our work to the existing literature. We then proceed to find an exact solution of the stationary state as well as the dynamics using an eigenfunction expansion method in Sec.~\ref{prelim}. 
Besides the exact results, we also provide accurate approximations for the frequency distribution when the selection coefficient is larger or smaller than the mutation rates. Sections \ref{stat} and \ref{dyn},  respectively, deal with these approximations in the stationary state and for the dynamics of the frequency distribution. A  discussion of the results follows in the concluding section.

\section{Model}
\label{model}

We consider an infinitely large asexual population of infinitely long diallelic sequences evolving in continuous time. All individuals carrying $k \geq 0$ deleterious mutations relative to the fittest individual are assumed to have the same (Malthusian) fitness $w_k=-s k, s \geq 0$ and said to belong to the fitness class $k$. We also assume a single-step mutation scheme in which a deleterious (beneficial) mutation increases (decreases) the fitness class by one and occurs at rate $U_d$ ($U_b$), but mutations to other classes are ignored. Then the population fraction $x_k(t)$ in the $k$th fitness class at time $t$ obeys the following differential-difference equations:
\begin{subequations}
\begin{align}
\dot{x}_0(t) &= U_b x_1(t)- U_d x_0(t) +s {\cal C}_1(t) x_0(t) ~,\label{x0eqn} \\
\dot{x}_k(t) &= U_b x_{k+1} (t)+ U_d x_{k-1}(t) - U x_k(t) - s (k-{\cal C}_1(t)) x_k(t)
~,~ k \geq 1 ~.\label{xkeqn}
\end{align}
\end{subequations}
In the above equations, $U=U_d+U_b$ is the total mutation rate and ${\cal C}_1(t)=\sum_{k=0}^\infty k ~x_k(t)$ is the average number of deleterious mutations in the population at time $t$. In the stationary state where the LHS of (\ref{x0eqn}) and (\ref{xkeqn}) is zero, we will denote the steady state fraction by $x_k$.  

It is easy to verify that the above set of equations respect the normalisation condition, 
\be
\sum_{k=0}^\infty x_k(t)=1 ~,~ t \geq 0 ~.
\label{norm1}
\ee
We also have the boundary condition 
\be
x_k(t) \stackrel{k \to \infty}{\longrightarrow 0} ~,
\label{bc1}
\ee 
which ensures that the total fraction remains finite at all times. To complete the definition of the model, we also need to specify the initial condition $x_k(0)$ for all $k$. The analysis in Sec.s \ref{prelim} and \ref{stat} holds for arbitrary initial conditions but in  Sec.~\ref{dyn}, we will assume that the population is initially localised in the fitness class $k^{(0)}$.  

The time evolution equations above for the frequency $x_k(t)$ are defined for $k \geq 0$ and therefore the maximum fitness is zero. However, (\ref{xkeqn}) has been used without an upper bound on fitness to describe the adaptation dynamics \cite{Tsimring:1996,Kessler:1997,Desai:2007a,Rouzine:2008,Brunet:2008,Park:2010}.
The latter is a reasonable model at short times for a population initially localised in a fitness class with many deleterious mutations  since the frequency in the fitness classes close to the fittest one can then be neglected. For this model, several works  \cite{Tsimring:1996,Kessler:1997,Desai:2007a,Park:2010} have shown that it does not have a traveling wave solution, and either a lower cutoff on the frequency modeling a finite population size \cite{Tsimring:1996} or discrete time dynamics \cite{Park:2010} are required to obtain it. In \cite{Rouzine:2008}, although a cutoff for the high-fitness edge is imposed in the deterministic model to account for the finite size of the population, this work also assumes a traveling wave solution for the continuous time model (see their (4) and (5)). However, as in \cite{Tsimring:1996,Kessler:1997,Desai:2007a,Park:2010}, our analysis of short time dynamics described in Sec.~\ref{genfn} also does not support a traveling wave behavior.


Although dynamics can be studied on an infinite line, a stationary state does not exist if the fitness is not bounded above. To see this, consider the steady state of  (\ref{xkeqn})  by setting the LHS to be zero. 
Since the frequency in any fitness class must not be negative and the first two terms on the RHS of  (\ref{xkeqn}) are positive, their contribution can be balanced if 
\be
k > {\cal C}_1-U/s=k_{\ast} ~,
\label{critmean}
\ee
thus leading to a maximum fitness corresponding to $-s k_{\ast}$. A previous analysis of (\ref{xkeqn}) with unbounded fitnesses finds a negative frequency distribution in the stationary state  and  claims that ``in the deterministic limit there is no true stationary state for arbitrary [beneficial mutation rate]..." (p. 1313, \cite{Goyal:2012}). However as discussed above, the loss of positivity is simply a consequence of the lack of upper bound on the fitness and in Sec.~\ref{stat}, we will show that the model defined by (\ref{x0eqn}) and (\ref{xkeqn}) has a nontrivial steady state. We also mention that if we set the maximum fitness to $-s k_\ast$  instead of zero, the frequency $x_m=0, m < k_\ast$ and $x_{m+k}, m \geq k_\ast$ is given by the solution $x_k$ of (\ref{x0eqn}) and (\ref{xkeqn}) \cite{Haigh:1978}.

Equations (\ref{x0eqn}) and (\ref{xkeqn}) are mathematically nontrivial for two reasons: first, they are nonlinear in the fractions $x_k(t)$ due to the last (selection) term on the RHS and second, they are second order difference equations 
in $k$ when both mutation rates are nonzero. 

\noindent(i) When the beneficial mutations are absent ($U_b=0$), in the stationary state, the boundary equation (\ref{x0eqn}) immediately yields the average number of deleterious mutations, ${\cal C}_1=U_d/s$. This result is very helpful since it renders (\ref{xkeqn}) to be  linear in the frequencies and we quickly arrive at the following well known result 
\citep{Kimura:1966,Haigh:1978}:
\be
{x}_k= \frac{e^{-U_d/s}}{k!} ~ \left(\frac{U_d}{s} \right)^k ~~~~~~ \left[U_b=0 \right] ~.
\label{Ub0}
\ee
The time-dependent frequency has also been obtained using a generating function method and shown to be Poisson-distributed \cite{Etheridge:2009}. 

\noindent(ii) When the deleterious mutations are absent ($U_d=0$), the stationary state is trivial ($x_k=\delta_{k,0}$). But  the short time dynamics can be obtained by extending the method of  \cite{Etheridge:2009} as  described in Sec.~\ref{genfn} (also, see \cite{Kessler:1997,Desai:2007a,Park:2010}).

\noindent(iii) In the neutral case ($s=0$), the nonlinear term on the RHS of  (\ref{x0eqn}) and (\ref{xkeqn}) vanishes. The stationary state frequency is then easily found to be  
\be
{x}_k = \left(1-\frac{U_d}{U_b} \right) ~ \left( \frac{U_d}{U_b} \right)^k ~~~~~~ \left[s=0, U_d < U_b \right] ~.
\ee
The condition $U_d < U_b$ arises due to the boundary condition (\ref{bc1}). 
However, in the parameter regime where $U_b < U_d$, the neutral population does not reach a steady state. In Appendix~\ref{neu}, we give the exact solution of the neutral dynamics in this parameter regime.

A brief summary of the results in the stationary state and for the dynamics is given in Table~\ref{summary}.

\begin{table}[t]
\begin{center}
\begin{tabular}{|c|c|c|c|c|}
\hline
& $U_b=0$   & $U_d=0$ & $s=0$ & all nonzero \\
\hline
 Stationary state &\cite{Kimura:1966,Haigh:1978} & trivial & none & Equation (\ref{xsel_ss}) \\
 Dynamics &\cite{Etheridge:2009} & \cite{Kessler:1997,Desai:2007a,Park:2010} & Equation (\ref{neu_exact}) & Equation (\ref{xktfinal2})\\
\hline
\end{tabular}
\caption{Summary of the results for the deterministic model defined by (\ref{x0eqn}) and (\ref{xkeqn}) where $U_d$ and $U_b$, respectively, denote deleterious and beneficial mutation rate and $s$ is the selection coefficient. In all the cases except when deleterious mutations are absent, it is assumed that $U_b < U_d$.}
\label{summary}
\end{center}
\end{table}

\section{Exact solution of the population frequency by eigenfunction expansion method}
\label{prelim}

We now proceed to find the population fraction $x_k(t)$ when all the three parameters, viz., $s, U_b, U_d$ are nonzero. In the following discussion, we assume that the mutation rate $U_d > U_b$ as in biologically realistic situations \cite{Perfeito:2007}. Since the equations (\ref{x0eqn}) and (\ref{xkeqn}) are nonlinear in the fractions $x_k(t)$, we work with the  unnormalised variables defined as \cite{Thompson:1974,Jain:2011c}
\be
z_k(t) = x_k(t) ~e^{-s \int_0^t dt' ~{\cal C}_1(t')}  ~,
\label{zkdefn}
\ee
which obey the following {\it linear} equations:
\begin{subequations}
\begin{align}
\dot{z}_0(t) &= U_b z_1 (t)- U_d z_0(t)  ~,\label{z_bdry} \\
\dot{z}_k(t) &= U_b z_{k+1}(t) + U_d z_{k-1} (t)- U z_k (t)- s k z_k(t)
~,~ k \geq 1 ~. \label{z_bulk}
\end{align}
\end{subequations}
Summing over $k$ on both sides of (\ref{zkdefn}) and using the normalisation condition (\ref{norm1}), we obtain the following relationship between the average ${\cal C}_1(t)$ and the  unnormalised frequencies $z_k(t)$:
\be
 \sum_{k=0}^\infty z_k(t) = e^{-s \int_0^t dt' ~{\cal C}_1(t')} ~.
 \label{zavg}
\ee
Using this in (\ref{zkdefn}), we immediately obtain
\be
x_k(t)= \frac{z_k(t)}{\sum_{m=0}^\infty z_m(t)} ~.
\ee

It is convenient to further define (p. 139, \cite{Kampen:1997})
\bea
y_k (t) &=& \left({\frac{U_b}{U_d}}\right)^{k/2} ~e^{(\sqrt{U_d}-\sqrt{U_b})^2 t} ~z_k(t) ~,\label{ykdefn} \\
\tau &=& t \sqrt{U_b U_d} ~,\\
\gamma &=& 2- \sqrt{{U_b}/{U_d}} ~,\\
S &=& \sqrt{s/U_b} \cdot \sqrt{s/U_d} ~.
\eea
In terms of these variables, we have 
\begin{subequations}
\begin{align}
\frac{\partial {y}_0 (\tau)}{\partial \tau} &= y_1(\tau)- \gamma y_0(\tau) ~, \label{y_bdry}  \\
\frac{\partial {y}_k (\tau)}{\partial \tau} &= y_{k+1}(\tau) + y_{k-1}(\tau) - (2+ S k) y_k(\tau) ~,~ k \geq 1  ~. \label{y_bulk}
\end{align}
\end{subequations}
The above set of equations involving two independent variables, viz., space and time can be solved by the eigenfunction expansion method (Chapter 5 and 6, \cite{Kampen:1997}). Since the differential operator $\partial/\partial \tau$  has eigenfunctions $e^{-\lambda \tau}$ with eigenvalues $-\lambda$, on expanding $y_k(\tau)$ as a linear combination of these eigenfunctions as 
\be
y_k(\tau)= \sum_{\lambda} c_\lambda e^{-\lambda \tau} \phi_k{(\lambda)} ~,~k \geq 0 ~,
\label{y_k2}
\ee
we obtain difference equations in one independent variable:
\begin{subequations}
\begin{align}
\phi_1-(\gamma-\lambda) \phi_0 &= 0  ~,\label{phi_bdry2} \\
\phi_{k+1}+\phi_{k-1}-(2+S k-\lambda) \phi_k &= 0 ~,~k \geq 1 ~.\label{phi_bulk} 
\end{align}
\end{subequations}
Equation (\ref{phi_bulk}) is an eigenvalue equation for a real symmetric matrix with eigenfunction $\phi_k$ and eigenvalue $-\lambda$. For such a matrix, it is possible to find a complete set of eigenvectors \cite{Courant:1953}. Moreover, the eigenvalues are real and the eigenfunctions corresponding to different eigenvalues are orthogonal and can be normalised to unity: 
\be
\sum_{k=0}^\infty \phi_k(\lambda) \phi_k(\lambda')=\delta_{\lambda,\lambda'} ~. 
\label{ortho}
\ee
The eigenvalues are determined by the boundary condition (\ref{phi_bdry2}) as explained below.  
The constants $c_\lambda$'s in (\ref{y_k2}) can be found using the initial condition and are given by 
\bea
c_\lambda &=& \sum_{m=0}^\infty \phi_m(\lambda) ~y_m(0) ~,\\
&=&  \sum_{m=0}^\infty \phi_m(\lambda) ~\left({\frac{U_b}{U_d}}\right)^{m/2} x_m(0) ~. 
\label{clam}
\eea
This can be seen by using (\ref{y_k2}) at $t=0$ and using the orthonormality condition (\ref{ortho}). 

Our remaining task now is to find the eigenfunctions $\phi_k$. We remark that if the fitness class $k$ is treated as a continuous variable, (\ref{phi_bulk}) reduces to a time-independent Schr{\"o}dinger equation for a particle in
a linear potential for which the eigenfunctions are known to be Airy function (and plane wave when $S$ is zero) \cite{Flugge:1974}. Here we are interested in finding the eigenfunctions in discrete fitness space with (Robin) boundary condition (\ref{phi_bdry2}). In the neutral case ($S=0$),  exact eigenfunctions and frequency $y_k(t)$  are obtained in Appendix~\ref{neu}. 
When the parameter $S$ is nonzero, the solution of (\ref{phi_bulk}) is a linear combination of the Bessel
function of first and second kind with order $\nu$ and argument $z$ that are denoted by $J_\nu(z)$ and $Y_\nu(z)$, respectively \cite{Ehrhardt:2004} (also, see Appendix~\ref{app_bess}):
\be
\phi_k(\lambda)= A(\lambda) J_{k+\frac{2-\lambda}{S}} \left( \frac{2}{S} \right)+ A'(\lambda) 
Y_{k+\frac{2-\lambda}{S}} \left( \frac{2}{S} \right) ~,~k \geq 0 ~.
\label{Bessel}
\ee
It is easy to check that (\ref{Bessel}) satisfies the eigenvalue equation (\ref{phi_bulk}) using the recurrence relation for the Bessel function ${\cal K}_\nu(z)$ given by (9.1.27, \cite{Abramowitz:1964}) 
\be
{\cal K}_{\nu-1}(z)+{\cal K}_{\nu+1}(z)= \frac{2 \nu}{z} {\cal K}_\nu(z) ~,
\label{recur} 
\ee
where ${\cal K}$ denotes $J, Y$. 
To find the constants $A, A'$, we invoke the boundary condition (\ref{bc1}) that the frequency $x_k(t) \to 0$ as the fitness class $k \to \infty$. From (\ref{zkdefn}) and (\ref{ykdefn}), it follows that $z_k(t)$ and $y_k(t)$  also obey this boundary condition. Using this large $k$ behavior,  we find that the coefficient $A'$ is zero since 
$Y_\nu(z)$ diverges for large $\nu$ (9.3.1, \cite{Abramowitz:1964}). Then using the orthonormality condition (\ref{ortho}), we find that 
\be
A^2 (\lambda)=\frac{1}{ \sum_{k=0}^\infty J^2_{k+\frac{2-\lambda}{S}} \left( \frac{2}{S} \right)} ~.
\label{a12}
\ee

The eigenvalues are determined using (\ref{Bessel}) in the boundary condition (\ref{phi_bdry2}) at $k=0$ and satisfy
\be
J_{1+\frac{2-\lambda}{S}} \left( \frac{2}{S}
\right)-(\gamma-\lambda)J_{\frac{2-\lambda}{S}} \left( \frac{2}{S}
\right) = 0  ~.
\label{phi_bdry}
\ee

Putting all the pieces together, we finally obtain
\bea
x_k(t) \propto \left({\frac{U_d}{U_b}}\right)^{k/2} \sum_{\lambda}  c_\lambda A(\lambda)    J_{k+\frac{2-\lambda}{S}} \left( \frac{2}{S} \right) e^{-\lambda \sqrt{U_b U_d} t} ~,
\label{xktfinal}
\eea
where $c_\lambda$ and $A(\lambda)$ are, respectively, given by (\ref{clam}) and (\ref{a12}), the eigenvalues by (\ref{phi_bdry}) and the proportionality constant is determined by the normalisation condition (\ref{norm1}). If $\lambda_\alpha, \alpha \geq 0$ denotes the $(\alpha+1)$th minimum eigenvalue, the above equation can be rewritten as 
\be
x_k(t)=\frac{\left({\frac{U_d}{U_b}}\right)^{k/2} \sum_{\alpha=0}^\infty  c_{\lambda_\alpha} A(\lambda_\alpha) ~ J_{k+\frac{2-\lambda_\alpha}{S}} \left( \frac{2}{S} \right) e^{-(\lambda_\alpha-\lambda_0) \sqrt{U_b U_d} t}}{\sum_{m=0}^\infty \left({\frac{U_d}{U_b}}\right)^{m/2} \sum_{\alpha=0}^\infty c_{\lambda_\alpha} A(\lambda_\alpha) ~  J_{m+\frac{2-\lambda_\alpha}{S}} \left( \frac{2}{S} \right) e^{-(\lambda_\alpha-\lambda_0) \sqrt{U_b U_d} t}} ~.
\label{xktfinal2}
\ee
This result can be verified by plugging it in (\ref{x0eqn}) and (\ref{xkeqn}) and using the relationship (\ref{zavg}) between the normalisation constant and the mean. 


\section{Stationary state frequency}
\label{stat}
 
To obtain the steady state, we take the limit $t \to \infty$ in (\ref{xktfinal2}) and find that only the minimum eigenvalue $\lambda_0$ contributes to the sum over the eigenvalues and the result is independent of the initial condition. We thus obtain the {\it exact}  stationary state frequency for an infinitely large population evolving under the joint action of deleterious and beneficial mutations and non-epistatic selection to be 
\be
x_k= \frac{\left(\frac{U_d}{U_b} \right)^{k/2} J_{k+\frac{2-\lambda_0}{S}} \left(\frac{2}{S}
  \right)}{\sum_{m=0}^\infty  \left(\frac{U_d}{U_b} \right)^{m/2}
  J_{m+\frac{2-\lambda_0}{S}} \left(\frac{2}{S}
  \right)} ~,
\label{xsel_ss}
\ee
where $\lambda_0$ is the minimum eigenvalue determined from  (\ref{phi_bdry}). Before proceeding further, we note  that the Bessel function $J_\nu(z)$ is an oscillatory function in both $\nu$ and $z$. However, since the eigenfunction corresponding to the minimum eigenvalue for a real symmetric matrix with homogeneous boundary condition cannot have zeros (p. 452, \cite{Courant:1953}),  the solution (\ref{xsel_ss}) satisfying (\ref{phi_bdry2}) and (\ref{phi_bulk})  is guaranteed to be positive for all $k \geq 0$. 

\begin{figure}
\includegraphics[width=1 \linewidth,angle=0]{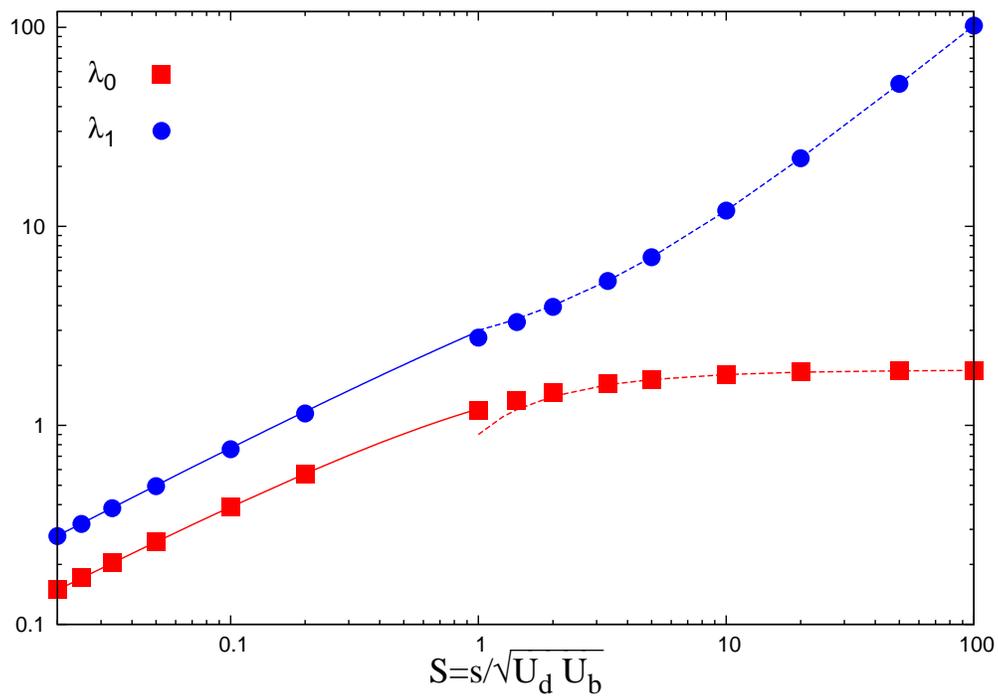}
\caption{Variation of the minimum and second minimum eigenvalue $\lambda_0$ and $\lambda_1$ with $S$. The points are  obtained by solving (\ref{phi_bdry}) numerically and the lines show the approximate expressions (\ref{smallS0}) and (\ref{largeS0}) for $\lambda_0$ and (\ref{lam1}) for $\lambda_1$ for $U_b=0.01 ~U_d, s=0.001$.}
\label{lam0_lam1_S}
\end{figure}

Using the above solution in (\ref{x0eqn}), we find that the average number of deleterious mutations in the steady state is given exactly by 

\bea
{\cal C}_1 &=& \frac{U_d}{s}- \frac{\gamma-\lambda_0}{S}  \label{kbarss1} 
\\
&=& \frac{U}{s}- \frac{2-\lambda_0}{S} \label{kbarss11}  ~.
\eea
As Fig.~\ref{lam0_lam1_S} shows, the minimum eigenvalue $\lambda_0$ initially increases with $S$ and approaches a constant asymptotically. Taking $S \to \infty$ in (\ref{phi_bdry}) and using that $J_0(0)=1, J_1(0)=0$ \cite{Abramowitz:1964}, we find that $\lambda_0 \to \gamma$ (also, see (\ref{largeS0}) below). 
Then, from (\ref{kbarss1}), it follows that beneficial mutations decrease the average number of deleterious mutations as also expected intuitively. Moreover, using the inequalities $\lambda_0 \leq \gamma \leq 2$ in (\ref{kbarss11}), it is easily  checked that the condition (\ref{critmean}) for the existence of the stationary state is satisfied.

The higher order cumulants such as variance and skewness can be found using a cumulant generating function as detailed in Appendix~\ref{app_cum}.  Alternatively, on multiplying both sides of (\ref{xkeqn}) in the steady state by $k$ and summing over $k$, we find the stationary state variance ${\cal C}_2={\overline {k^2}}-{\overline k}^2$ to be 
\be
{\cal C}_2= \frac{U_d}{s}-\frac{U_b}{s} (1-x_0) ~,
\label{exactvar}
\ee
which shows that  beneficial mutations decrease the width of the distribution also. Furthermore, as the inset of Fig.~\ref{kbar_S_1} shows, the variance to mean ratio is greater than one and therefore the frequency distribution is non-Poissonian  when $U_b$ is nonzero. 

To obtain some insight into the behavior of the equilibrium frequency given by (\ref{xsel_ss}) above, we now consider two parameter regimes where the ratio $S$ of the selection coefficient to the mutation rates is large or small relative to one. 


\subsection{When the parameter $S$ is large} 
\label{sub_largeS}

The parameter $S \gg 1$ when (i) $U_b < s$ and (ii) either  $U_d < s$ or $s < U_d < s^2/U_b$. When $U_b=0$, as (\ref{Ub0}) shows, the average number of deleterious mutations in the stationary state equals $U_d/s$. Therefore, when $U_b$ is turned on, we expect that beneficial mutations do not have a significant effect when $U_d < s$ but they can decrease the mean ${\cal C}_1$ substantially when $U_d > s$. The analysis given below is in agreement with these expectations. 

As described in Appendix~\ref{app_largeS} and shown in Fig.~\ref{lam0_lam1_S}, the minimum eigenvalue $\lambda_0$ when $S$ is large is given by
\bea
\lambda_0 \approx \gamma- S^{-1} ~.
\label{largeS0} 
\eea
On plugging  (\ref{largeS0}) in (\ref{xsel_ss}), we obtain
\bea
x_k \propto \left(\frac{U_d}{U_b} \right)^{k/2}  J_{k+\frac{U_b}{s} (1+\frac{U_d}{s})} \left(\frac{2}{S} \right) ~,
\eea
which, on using  the series representation (\ref{series}) of Bessel function, yields 
\be
x_k \propto  \left( \frac{U_d}{s} \right)^k  \sum_{m=0}^\infty \left(\frac{U_b U_d}{s^2} \right)^m
\frac{(-1)^m}{m ! (m+k+\frac{U_b}{s} (1+\frac{U_d}{s}))!} ~.
\label{largeSxk}
\ee
We check that the above solution reduces to (\ref{Ub0}) when $U_b=0$. To see the effect of beneficial mutations, it is useful to expand (\ref{largeSxk}) in a power series in $U_b/s$ as was done recently in \cite{James:2016} and described here briefly in Appendix~\ref{app_largeS2}. This discussion as also Fig.~\ref{xklargeS} show that a nonzero $U_b$ has a significant effect when $U_d > s$.

Furthermore, using (\ref{largeS0})  in the exact expression (\ref{kbarss1}) for the average ${\cal C}_1$, we find that \cite{James:2016}
\be
{\cal C}_1 \approx \frac{U_d}{s} \left(1- \frac{U_b}{s} \right) ~.
\label{kbarSl}
\ee
Since $U_b/s$ is small for large $S$, using (\ref{Ub0}) for the frequency $x_0$ in (\ref{exactvar}), we find that the variance is well approximated by
\bea
{\cal C}_2 &\approx& \frac{U_d}{s}-\frac{U_b}{s} (1-e^{-U_d/s}) ~,\label{varSl} \\
&=&
\begin{cases}
{\cal C}_1 & ~,~U_d \ll s ~, \\
\frac{U_d-U_b}{s} & ~,~U_d \gg s ~.
\end{cases}
\eea
Thus the variance is close to mean (\ref{kbarSl}) when $U_d/s \ll 1$ but larger in the opposite parameter regime. The above approximations are tested against the corresponding exact results in Fig.~\ref{kbar_S_1} and we see a good agreement. 


\subsection{When the parameter $S$ is small} 
\label{sub_smallS}

The parameter $S \ll 1$ when (i) $s < U_d$ and (ii) either $s < U_b$ or $U_b < s < \sqrt{U_b U_d}$. A biologically relevant situation where $S$ can be small arises in the case of mutators where mutation rates can be as high as $10^{-2}$ \cite{Wielgoss:2013}. Then for selection coefficient in the range $10^{-4}-10^{-3}$, one obtains $S \sim 0.01-0.1$. 

\begin{figure}
\includegraphics[width=1 \linewidth,angle=0]{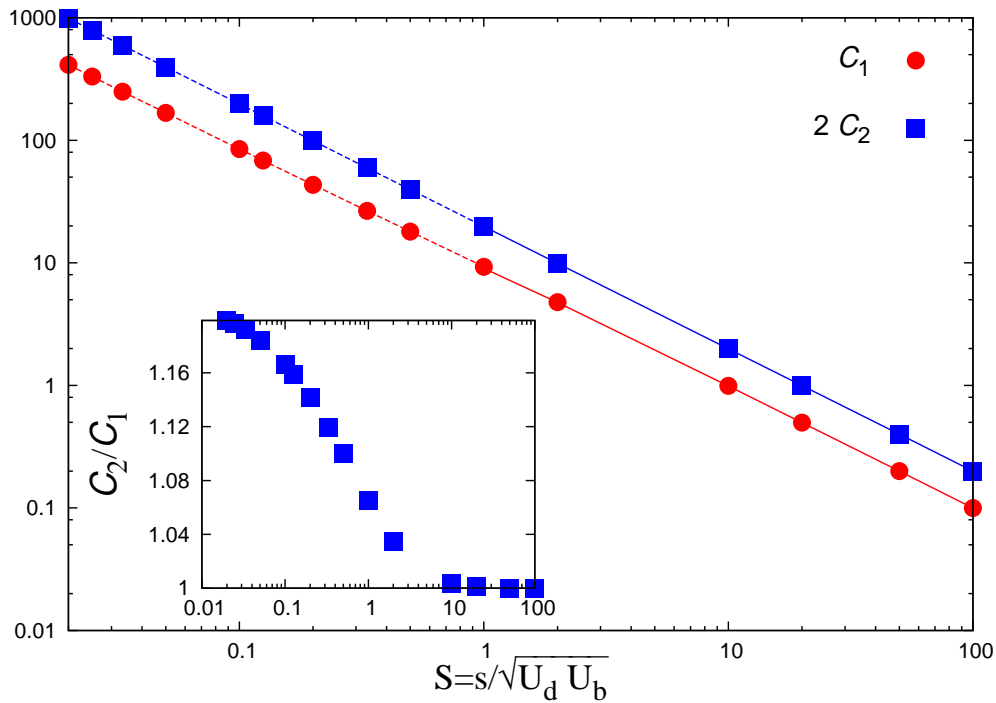}
\caption{Main figure shows the exact mean ${\cal C}_1$ and variance ${\cal C}_2$  in the stationary state given, respectively, by (\ref{kbarss1}) and (\ref{exactvar}) for $U_b=0.01 ~U_d, s=0.001$. The lines show the approximate expressions (\ref{kbarSl}), (\ref{meanS}) for mean and (\ref{varSl}), (\ref{varS}) for variance. The variance is scaled by a factor $2$ for clarity. The exact ratio ${\cal C}_2/{\cal C}_1$ shown in the inset supports the non-Poissonian nature of the frequency distribution in the steady state.}
\label{kbar_S_1}
\end{figure}

\begin{figure}
\includegraphics[width=1 \linewidth,angle=0]{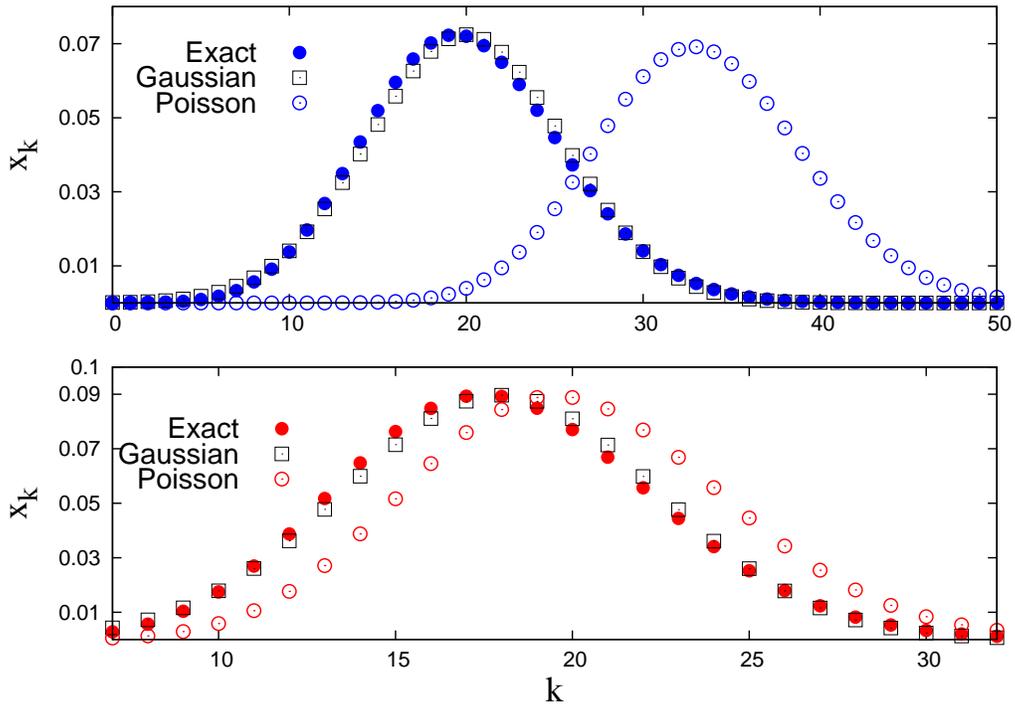}
\caption{Steady state distribution when $s \ll \sqrt{U_b U_d}$: Comparison of the exact distribution (\ref{xsel_ss}), Gaussian approximation (\ref{xksSapprox2}) and Poisson distribution (\ref{Ub0}). The parameters in the top and bottom panel are $s=0.003, U_b=0.009, U_d=0.1$ and $U_b=0.001, s=0.005, U_d=0.1$, respectively.}
\label{xksmallS}
\end{figure}

For small $S$, the minimum eigenvalue shown in Fig.~\ref{lam0_lam1_S} is calculated in Appendix~\ref{app_smallS} and given by 
\bea
\lambda_0 = \left( \frac{9 \pi}{8} \right)^{2/3}~S^{2/3} - \frac{S}{\gamma-1} ~.
\label{smallS0}
\eea 
Using this in (\ref{kbarss1}), we find that 
\bea
{\cal C}_1 
\approx \frac{U_d-U_b}{s}- \left[ \frac{2}{S} -\left(\frac{9 \pi}{8} \right)^{2/3} \frac{1}{S^{1/3}} +\frac{1}{\gamma-1} \right] ~.
\label{meanS}
\eea
As our numerical analysis of (\ref{xsel_ss}) shows that the fraction $x_0 \sim e^{-1/S}$ for small $S$ (also, see Fig.~\ref{xksmallS}), the variance (\ref{exactvar}) can be approximated by
\be
{\cal C}_2 \approx \frac{U_d-U_b}{s}  ~,
\label{varS}
\ee
as also seen in (\ref{varSl}) when $s < U_d$. The above equation  also shows that the variance is larger than the mean as illustrated in Fig.~\ref{kbar_S_1}. The above approximations are in good agreement with the exact results, see Fig.~\ref{kbar_S_1}.  

An analysis of the frequency distribution (\ref{xsel_ss}) for small $S$  described in Appendix~\ref{app_smallS2} suggests a Gaussian approximation for the frequency distribution,  
\be
x_k \approx \sqrt{\frac{1}{2 \pi  {\cal C}_2}} \exp \left[- \frac{(k -{\cal C}_1)^2}{2 {\cal C}_2}  \right] ~,
\label{xksSapprox2}
\ee
where the mean and variance are given, respectively, by (\ref{meanS}) and (\ref{varS}). Figure~\ref{xksmallS} compares the above approximation with the exact distribution (\ref{xsel_ss}) and we see a quite good agreement. However, it should be noted that unlike (\ref{xksSapprox2}), 
 the exact frequency distribution is not symmetric about the mean. The skewness defined as ${\cal C}_3/{\cal C}_2^{3/2}$,  where ${\cal C}_3$ is the third cumulant, is a measure of the asymmetry of the distribution. Due to (\ref{c3cum}) in the stationary state, on neglecting the frequency $x_0$, we find a nonzero ${\cal C}_3=U/s$. Figure~\ref{xksmallS} also shows that  both mean and variance are considerably affected by beneficial mutations when $s < U_b < U_d$.

\section{Dynamics of the population frequency}
\label{dyn}

\begin{figure}
\includegraphics[width=1 \linewidth,angle=0]{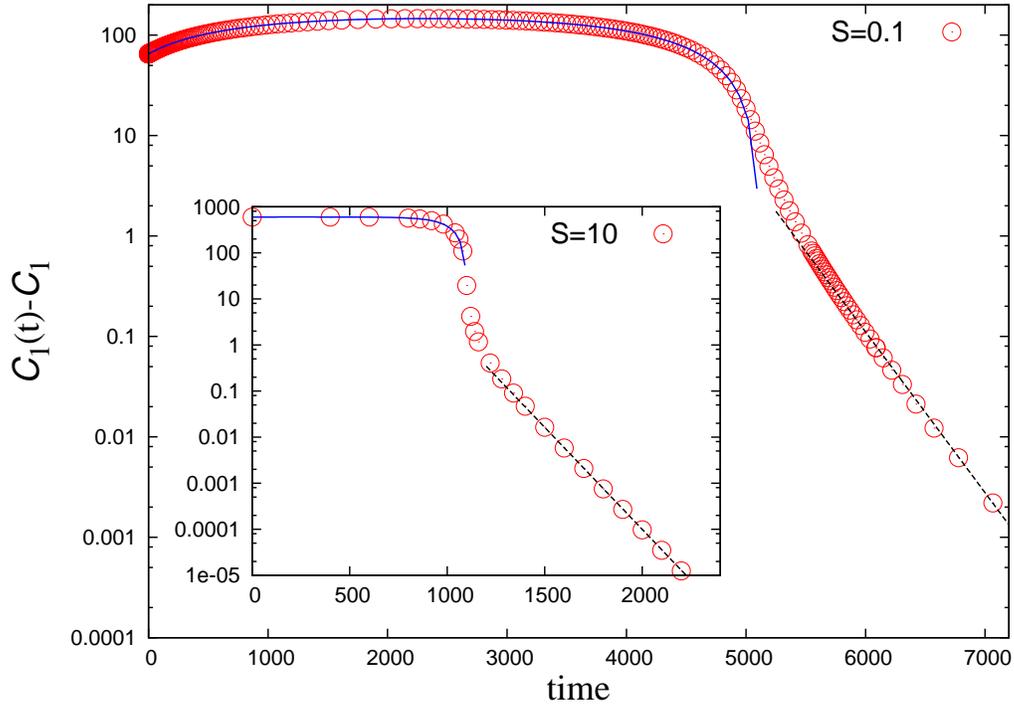}
\caption{Dynamics of the mean deviation from the stationary state, ${\cal C}_1(t)-{\cal C}_1$ for two values of $S=s/\sqrt{U_b U_d}$ with initial population located in the fitness class $k^{(0)}$. The exact dynamics obtained by numerically integrating (\ref{x0eqn}) and (\ref{xkeqn}) are shown by points while the solid (blue) lines show the short time dynamics (\ref{shorttime}) and the broken (black) line shows the relaxation dynamics $b e^{-R t}$, where the relaxation rate $R$ is given by (\ref{relax}). The parameters in the main and inset are $U_b=s=0.001, U_d=0.1, k^{(0)}=150, b=4.515 \times 10^{8}, {\cal C}_1=84.8958$ and $U_b=0.0001, s=U_d=0.01, k^{(0)}=600, b=71561, {\cal C}_1=0.990147$, respectively.}
\label{kbar_t}
\end{figure}

In the last section, we discussed the stationary state and now turn to the dynamics of the frequency distribution. We will focus on the time dependence of the average ${\cal C}_1(t)$ starting from a monomorphic initial condition given by
\be
x_k(0)=\delta_{k,k^{(0)}}  ~.
\label{ic1}
\ee
As the exact expression (\ref{xktfinal2}) for the time-dependent frequency involves a sum over a large number of eigenvalues, the dynamics are more efficiently studied by solving the differential equations (\ref{x0eqn}) and (\ref{xkeqn}) numerically. The results for the mean thus obtained are shown in Fig.~\ref{kbar_t} for two values of $S$ 
and, we observe (i) a short time regime where the population is far from the stationary state, (ii) an intermediate time regime where the mean changes quickly and (iii) a long time relaxation regime where the population is close to the steady state.

The initially monomorphic population first spreads over the genotypic space due to mutations followed by an increase in the frequency of high fitness genotypes as a result of selection. At short enough times, one can understand the dynamics away from the stationary state by ignoring the boundary at $k=0$ (see, Sec.~\ref{genfn} below). Note that this holds even if the population is initially located in the zeroth fitness class since, as Fig.~\ref{kbar_t} shows, the mean initially increases because $U_d > U_b$. 
However once the population is close to the stationary state (i.e., ${\cal C}_1(t)- {\cal C}_1 \lesssim 1$ in Fig.~\ref{kbar_t}), the boundary at the zeroth fitness class becomes important. As discussed in Sec.~\ref{relax2}, at long enough times, it is sufficient to retain the minimum and second minimum eigenvalue in the sum over the eigenvalues in (\ref{xktfinal2}) to determine the time to relax to the stationary state.

\subsection{Dynamics far from the stationary state}
\label{genfn}

The dynamics of the $n$th cumulant ${\cal C}_n(t)$ are described in Appendix~\ref{app_cum}. 
For the initial condition (\ref{ic1}), at short enough times, we can set the frequency in the fittest class to be approximately zero in (\ref{gencum}) to obtain
\be
{\dot {\vec {\cal C}}}(t) = -s {\hat D} {\vec {\cal C}}(t)  + {\vec U} ~, 
\label{Cntdiff}
\ee
where ${\vec {\cal C}}$ and $\vec U$ are column vectors whose $n$th element is given by ${\cal C}_n$ and $U_d+ (-1)^n U_b$ respectively and  $\hat D$ is an upper shift matrix with matrix element $D_{mn}=\delta_{m+1,n}$. The above equation can be straightforwardly solved for arbitrary initial condition and for (\ref{ic1}), we obtain
\be
{\cal C}_n(t)= \frac{U_n}{s}  \sinh (s t)- \frac{U_{n+1}}{s} (\cosh(s t)-1)+ k^{(0)} \delta_{n,1} ~.
\label{Cnt}
\ee
Using (\ref{Cnt}) in (\ref{cumdyn}), the generating function of the population fraction, $F(\xi,t)= \sum_{k=0}^\infty x_k(t) e^{-\xi k}$ can also be obtained and given by
\be
\ln F(\xi,t)= -k^{(0)} \xi - \frac{U_d}{s} (1-e^{-s t}) (1-e^{-\xi}) - \frac{U_b}{s} (1-e^{s t}) (e^{\xi}-1) ~.
\ee
The above result generalises (53) of \cite{Kessler:1997} who obtained it for special values of the parameters. 

Due to (\ref{Cnt}), the approximate short time dynamics of mean ${\cal C}_1(t)$ and variance ${\cal C}_2(t)$ are given by
\bea
{\cal C}_1(t) &=& \frac{U_d}{s} (1-e^{-s t}) + \frac{U_b}{s} (1-e^{s t})+ k^{(0)} ~, \label{shorttime} \\
{\cal C}_2(t) &=& \frac{U_d}{s} (1-e^{-s t}) + \frac{U_b}{s} (e^{s t}-1) ~.
\eea
The above equations show that for $t \ll 1/s$, both mean and variance change linearly with time with a slope that depends only on the mutation rates but for longer times, the mean varies exponentially fast at a rate $s^{-1}$. The short time dynamics of the mean given by (\ref{shorttime}) are valid as long as $|{\cal C}_1(t)- {\cal C}_1|$ is large and agree with the numerical results shown in Fig.~\ref{kbar_t}.

\subsection{Dynamics close to the stationary state}
\label{relax2}

When beneficial mutations are absent, the frequency $x_k(t)$ is  Poisson-distributed with mean $(U_d/s) (1-e^{-s t})$  \cite{Maia:2003,Etheridge:2009} and thus approaches the stationary state at rate $s$, independent of the deleterious mutation rate. When the beneficial mutation rate $U_b$ is nonzero, the dynamical evolution of the frequency is given by (\ref{xktfinal2}). At large but finite times,  it is a good approximation to retain only the terms containing the minimum and second minimum  eigenvalues in the sum over the eigenvalues in  (\ref{xktfinal2}).  Thus the frequency $x_k(t)$ relaxes to the steady state exponentially fast at rate 
\be
R=(\lambda_1-\lambda_0) \sqrt{U_b U_d} ~,
\label{relax}
\ee
where $\lambda_1$ is the second minimum eigenvalue. Unlike $\lambda_0$, the second minimum eigenvalue $\lambda_1$ is an increasing function of $S$ as shown in Fig.~\ref{lam0_lam1_S}. 

The second minimum eigenvalue is calculated in Appendix~\ref{app_largeS} and \ref{app_smallS} and given by 
\be
\lambda_1 =
\begin{cases}
\left( \frac{21 \pi}{8}\right)^{2/3} S^{2/3} -\frac{S}{\gamma-1}   &~,~S \ll 1 ~,\\
S + 2 & ~,~S \gg 1 ~.
\end{cases}  
\label{lam1}
\ee
This yields the relaxation rate $R=s+U_b$ for large $S$  which shows that the population reaches the stationary distribution faster in the presence of beneficial mutations. For small $S$, we get $R \sim s^{2/3} (U_b U_d)^{1/6}$ which approaches zero as $s \to 0$ in accordance with the neutral case where the population never reaches a stationary state, see Appendix ~\ref{neu}.  The relaxation dynamics of the mean ${\cal C}_1(t)$ are in agreement with the above results as shown in Fig.~\ref{kbar_t}. 

On comparing the results for short and long time dynamics, we find that while the former occurs over a time scale $\sim s^{-1}$ independent of the mutation rates, the relaxation time is determined by both selection and mutation. 
\section{Conclusions}
\label{concl}

In this article, we presented the exact solution (\ref{xktfinal2}) for the frequency distribution at all times for the model defined by (\ref{x0eqn}) and (\ref{xkeqn}). Our results summarised in Table~\ref{summary} generalise the earlier ones in \cite{Kimura:1966,Haigh:1978,Etheridge:2009} 
by including beneficial mutations and extend the treatment in  \cite{Tsimring:1996,Kessler:1997,Desai:2007a,Rouzine:2008,Brunet:2008,Park:2010}
to all times including the stationary state limit. We discussed the biologically realistic situation where the beneficial mutation rate is smaller than its deleterious counterpart \cite{Perfeito:2007} but, for completeness, we explore 
the parameter regime $U_d \ll U_b$ in Appendix~\ref{app_Udsmall}.

Here we considered a mutation scheme in which the mutation rate per sequence is same for all sequences, irrespective of their fitness. 
The evolution of an infinitely large population on additive fitness landscapes when the mutation rates depend  linearly on the number of loci carrying deleterious allele has also been studied \cite{Woodcock:1996,John:2015} and the relationship of this mutation scheme with the one studied in this article is elucidated in Appendix~\ref{app_reln}.
In the fitness-dependent mutation rate model  \cite{Woodcock:1996,John:2015}, when the number of loci carrying the  deleterious allele is small, the beneficial mutation rate vanishes in the limit of infinitely long sequence. This has the immediate consequence that the stationary state properties are not affected by beneficial mutations in this mutation scheme \cite{John:2015}.

In contrast, for the model studied here, the general effect of beneficial mutations is to decrease both mean and variance in the stationary state 
(see, (\ref{kbarss1}) and (\ref{exactvar})) but the extent to which this happens depends on the strength of selection relative to mutations. We find that

(i) when $U_b <  U_d < s$, beneficial mutations have a minor effect since the mean number of deleterious mutations in the absence of beneficial mutations is already close to zero, 

(ii) when $U_b < s < U_d$, beneficial mutations decrease the mutational load significantly and the frequency is enhanced (diminished) in fitness classes below (above) $U_d/s$ and 

(iii) when $s < U_b <  U_d$, both  mean and  variance decrease considerably and the frequency distribution is a nontrivial function.

Here we have focused on the deterministic evolution and ignored the effect of random genetic drift. However  in a finite population, the bulk of the distribution is expected to be well described by deterministic distributions \cite{Desai:2007a,Rouzine:2008,Brunet:2008,Goyal:2012}. A detailed study of the evolution of finite populations exploiting the results presented here will be taken up in future. \\

\noindent Acknowledgements: The authors thank Arul Lakshminarayan and three reviewers for several helpful comments.

\clearpage 


\appendix

\setcounter{section}{0}
\setcounter{equation}{0}
\makeatletter 
\renewcommand{\thesection}{A\@arabic\c@section} 
\renewcommand{\theequation}{A\@arabic\c@equation} 
\bigskip


\section{Neutral dynamics using eigenfunction expansion method}
\label{neu}

The treatment below for the unnormalised frequency $y_k(t)$ essentially follows Chap. 7, \cite{Kampen:1997} and here we briefly describe our results.  
For $S=0$, the solution of the eigenvalue equation (\ref{phi_bulk}) is given by 
\be
\phi_k=C_+ a_+^k +C_- a_-^k~,~ k \geq 0 ~,
\ee
where $a_\pm$ are solutions of the quadratic equation $a^2+ (\lambda-2) a+1=0$ and the coefficients $C_+$ and $C_-$ are related due to the boundary equation (\ref{phi_bdry2}). 
Since $a_+ a_-=1$, it is convenient to write $a_{\pm}=e^{\pm i q}$ where $q$ is real. The latter condition is required to ensure that the eigenfunction $\phi_k$ does not diverge at large $k$ (see (\ref{bc1})).  
Using $a_+ + a_-= 2 -\lambda$, we find that the eigenvalues form a continuous spectrum and are given by 
\be
\lambda =4 \sin^2 \left(\frac{q}{2} \right) ~,~ 0 \leq q \leq \pi ~.
\label{lamneu}
\ee
Since the ratio $C_+/C_-$ determined using (\ref{phi_bdry}) has unit modulus, we can write 
\be
\frac{C_+}{C_-}= - \frac{\gamma-2+e^{i q}}{\gamma-2+e^{-i q}} =e^{i 2 \eta(q)} ~,
\label{etadefn}
\ee
and finally arrive at 
\be
\phi_k(q)= \sqrt{\frac{2}{\pi}} \cos(q k+\eta(q)) ~,
\label{eiegnneu}
\ee
where we have used the orthonormality condition (\ref{ortho}) to determine the proportionality constant.  

For the initial condition $x_k(0)=\delta_{k,0}$, on replacing the sums in  (\ref{y_k2}) and (\ref{clam}) by integrals  (as the eigenvalues are continuous), we obtain
\bea
y_k(t) &=& \int_0^\pi \phi_k(q) \phi_0(q) e^{-2 (1-\cos q) \sqrt{U_b U_d} t} dq ~,\\
&=& \frac{2}{\pi} \int_0^\pi dq \sin q 
\frac{(\gamma-2) \sin (q k) +\sin (q k+q)}{(\gamma-2)^2+2 (\gamma-2) \cos q+1} ~e^{-2 (1-\cos q) \sqrt{U_b U_d} t} ~,
\label{neu_exact}
\eea
where the last expression follows on using (\ref{etadefn}) in (\ref{eiegnneu}). The above integral does not appear to be exactly solvable, but in the scaling limit $q \to 0, k, t\to \infty$ with $q^2 t$ and $q k$ finite, the above equation simplifies to give 
\bea
y_k (t) &\approx& \frac{2}{\pi} \int_0^\infty dq  \frac{(\gamma-1) q \sin (q k) + q^2 \cos (q k)}{(\gamma-1)^2} e^{-q^2 \sqrt{U_b U_d} t} ~, \\
&=&  \frac{1}{2 \sqrt{\pi}} ~\frac{k}{\gamma-1} ~\frac{e^{-\frac{k^2}{4 t \sqrt{U_b U_d}}}}{(t \sqrt{U_b U_d})^{3/2}} ~.
\eea
\section{Some properties of the Bessel functions}
\label{app_bess}

If ${\cal K}$ denotes $J, Y$, the Bessel function ${\cal K}_\nu(z)$ is defined as the solution of the following differential equation (9.1.1,\cite{Abramowitz:1964})
\be
z^2 \frac{d^2 {\cal K}_\nu(z)}{d z^2} +z \frac{d {\cal K}_\nu(z)}{d z} + (z^2-\nu^2) {\cal K}_\nu(z)=0 ~.
\ee
The Bessel function of the second kind $Y_\nu(z)$ is related to the Bessel function of the first kind $J_\nu(z)$ by (9.1.2,\cite{Abramowitz:1964})
\be
Y_\nu(z)= \frac{\cos(\nu \pi) J_\nu(z)-J_{-\nu}(z)}{\sin (\nu \pi)} ~.
\ee
For the series representation of $J_\nu(z)$, see (\ref{series}) below; the asymptotic expansions of $J_\nu(z)$ for $z > \nu$ and $z < \nu$ are given in  (\ref{Jsec}) and (\ref{asymJ}), respectively.

\section{Cumulants of the number of deleterious mutations}
\label{app_cum}

Following \cite{Etheridge:2009}, we first define the generating function of the frequency as 
\be
F(\xi,t)= \sum_{k=0}^\infty x_k(t) e^{-\xi k} ~.
\ee
Multiplying (\ref{x0eqn}) and (\ref{xkeqn}) by $e^{-\xi k}$ and summing over $k$, we obtain
\be
\frac{d \ln F(\xi,t)}{dt}=s {\cal C}_1(t)+s \frac{d \ln F(\xi,t)}{d \xi}- U_d (1-e^{-\xi}) +U_b (e^{\xi}-1) \left(1-\frac{x_0}{F(\xi,t)} \right)~.
\label{cumdynn}
\ee

The $n$th cumulant ${\cal C}_n(t), n=1,2, ...$ of the number of deleterious mutations is related to $F(\xi,t)$ through
\be
\ln F(\xi,t)= \sum_{n=1}^\infty {\cal C}_n(t) \frac{(-\xi)^n}{n!} ~.
\label{cumdyn}
\ee
Using the above equation in (\ref{cumdynn}), we get 
\bea
\sum_{n=1}^\infty {\dot {\cal C}}_n(t) \frac{(-\xi)^n}{n!} &=& -s \sum_{n=1}^\infty {{\cal C}}_{n+1}(t) \frac{(-\xi)^n}{n!} +U_d \sum_{n=1}^\infty \frac{(-\xi)^n}{n!} \nonumber  \\
&+& U_b \left(\sum_{n=1}^\infty \frac{\xi^n}{n!} \right) ~\left(1- \frac{x_0(t)}{F(\xi,t)} \right) ~.
\label{gencum}
\eea
Matching the coefficient of $\xi^n$ for $n=1, 2$ on both sides, we find that 
\bea
{\dot {\cal C}}_1(t) &=& -s {{\cal C}}_{2}(t) +U_d -U_b  (1-x_0(t)) ~, \label{varcum}\\
 {\dot {\cal C}}_2(t) &=& -s {{\cal C}}_{3}(t) +U - U_b x_0(t) (1+2 {\cal C}_1(t)) ~.\label{c3cum}
 \eea


\section{Approximate expressions for the eigenvalues when $S$ is large}
\label{app_largeS}

The Bessel function of the first kind has the following series representation (9.1.10, \cite{Abramowitz:1964}):
\be
J_\nu(z)= \sum_{m=0}^\infty \frac{(-1)^m}{m! (\nu+m)!} ~\left( \frac{z}{2} \right)^{2 m+\nu} ~.
\label{series}
\ee
For large $S$, keeping the first two terms in the above series and using it 
in the eigenvalue equation (\ref{phi_bdry}), we get 
\be
(\lambda-\gamma) (\lambda-S-2) \approx 1 ~.
\ee
Solving the above quadratic equation, we obtain the first two eigenvalues $\lambda_0$ and $\lambda_1$ given in (\ref{largeS0}) and (\ref{lam1}) respectively. Figure~\ref{lam0_lam1_S} shows a comparison between the exact eigenvalues  obtained numerically using (\ref{phi_bdry}) and the above approximations. Our numerical analysis of (\ref{phi_bdry}) also suggests that the $\alpha$th eigenvalue is given by 
\be
\lambda_\alpha \approx \alpha S+2   ~,~ \alpha=1, 2, ... ~.
\ee

\section{Approximate expression for the stationary distribution when $S$ is large}
\label{app_largeS2}

Noting that the terms corresponding to $m=0, 1$ in the summand on the RHS of (\ref{largeSxk}) contribute to the leading order term in $U_b/s$, we obtain
\bea
x_k \propto \frac{\mu^k}{k!}  \left[1+ \frac{U_b}{s} \left( (\gamma_{EM}-H_k) (1+\mu) - \frac{\mu}{k+1} \right) \right] ~,
\label{app22}
\eea
where $\mu=U_d/s$. In the above equation, we have used that $k!/(k+\epsilon)! \approx 1+ \epsilon (\gamma_{EM}-H_k)$ for small $\epsilon$ where $\gamma_{EM} \approx 0.577...$ and $H_k=\sum_{i=1}^k i^{-1}$ is the Harmonic number. On fixing the proportionality constant using (\ref{norm1}), we recover (A7) of \cite{James:2016}:
\be
x_k=  \frac{e^{-\mu}\mu^k}{k!} \left[ 1+ \frac{U_b}{s} \left( \sum_{m=0}^\infty \frac{e^{-\mu}\mu^m}{m!}  (H_m (1+\mu)+ \frac{\mu}{m+1}) -H_k (1+\mu) -\frac{\mu}{k+1} \right)\right]~.
\label{app222}
\ee
For $U_d < s$, the above expression shows that the distribution is close to the Poisson distribution (\ref{Ub0}). However, for $U_d > s$, we obtain \cite{James:2016}
\be
x_k \approx \frac{e^{-U_d/s}}{k!} \left(\frac{U_d}{s} \right)^k \left[1+\frac{U_b U_d}{s^2} \ln \left(\frac{U_d}{s k} \right) \right] ~,
\ee
on using $\sum_{m=0}^\infty z^m H_m/m! \approx e^z \ln z$ for large $z$ and $H_k \approx \ln k + \gamma_{EM}$ for large $k$ in (\ref{app222}). This result shows that the beneficial mutations enhance the frequency in fitness classes $k < U_d/s$ and diminish it in higher ones (also, see Fig.~\ref{xklargeS}). 


\section{Approximate expressions for the eigenvalues when $S$ is small}
\label{app_smallS}

For the Bessel function $J_\nu(z), z > \nu$, the asymptotic expansion for large orders is given by (9.3.3, \cite{Abramowitz:1964})
\be
J_\nu(\nu \sec \beta) \sim \frac{\cos \left( \nu (\tan \beta-\beta)-(\pi/4) \right)}{\sqrt{(\nu/2) \pi \tan \beta}} ~,~0 < \beta < \pi/2 ~.
\label{Jsec}
\ee
As shown in Fig.~\ref{lam0_lam1_S}, the eigenvalues $\lambda_0, \lambda_1$ are an increasing function of $S$ and approach zero as $S \to 0$ (also, see (\ref{lamneu}) for the neutral case). Then using (\ref{Jsec}) in (\ref{phi_bdry}) and carrying out a small $\lambda$ expansion, we obtain 
\be
\frac{\cos (\frac{2 \lambda^{3/2}}{3 S}-\frac{\pi}{4} -\sqrt{\lambda})}{\cos (\frac{2 \lambda^{3/2}}{3 S} -\frac{\pi}{4})}=\gamma-\lambda ~.
\label{lamsmm}
\ee
After some algebra, the above simplifies to 
\be
\tan \left( \frac{2}{3} \frac{\lambda^{3/2}}{S} \right) = - \frac{\gamma-1+\sqrt{\lambda}-\lambda}{\gamma-1-\sqrt{\lambda}-\lambda} ~.
\label{taneqn}
\ee
The above equation immediately suggests that  the eigenvalue $ \lambda \sim S^{2/3}$ so that the  RHS can be nonzero and finite. Guided by this observation and a numerical analysis of (\ref{phi_bdry}), we expect that the $\alpha$th eigenvalue is of the following form:
\be
\lambda_\alpha = \lambda_\alpha^{(0)} S^{2/3} +  \lambda_\alpha^{(1)} S ~.
\label{lamL1}
\ee
Substituting this in (\ref{taneqn}) and expanding both sides of the equation for small $S$, we find that
\bea
\lambda_\alpha^{(0)}&=&  \left( \frac{3 \pi (4 \alpha+3)}{8}\right)^{2/3} ~,\\
\lambda_\alpha^{(1)} &=& -\frac{1}{\gamma-1} ~.
\label{lamL2}
\eea
As we have assumed $\lambda$ to be small to arrive at (\ref{lamsmm}), the above results for the eigenvalues are valid for small $\alpha$. For larger $\alpha$, our numerical analysis of (\ref{phi_bdry}) suggests that the eigenvalues increase linearly with $S$.
\section{Approximate expression for the stationary distribution when $S$ is small}
\label{app_smallS2}

The asymptotic expansion of the Bessel function $J_\nu(z), z < \nu$ for large orders is given by (9.3.1, \cite{Abramowitz:1964})
\be
J_\nu(z) \sim \frac{1}{\sqrt{2 \pi \nu}} \left(\frac{z e}{2 \nu} \right)^\nu ~.
\label{asymJ}
\ee
Using this in (\ref{xsel_ss}), we obtain  the steady state frequency to be 
\be
x_k \propto \left( \sqrt{\frac{U_d}{U_b}} \right)^{k+\delta} \frac{1}{\sqrt{k+\delta}} \left( \frac{e/S}{k+\delta} \right)^{k+\delta} ~,
\ee
where $\delta=(2-\lambda_0)/S$ and $\lambda_0$ is given by (\ref{smallS0}). A Gaussian approximation for the above expression can be obtained by writing $x_k \propto e^{I(k)}$ and expanding $I(k)$ about its turning point ${\tilde k}=(U_d/s)-\delta$ up to quadratic orders in the deviation $k-{\tilde k}$. On fixing the normalisation, we obtain 
\be
x_k \approx \sqrt{\frac{s}{2 \pi  U_d}} \exp \left[- \frac{s}{2 U_d}  \left(k - \frac{U_d}{s} + \frac{2-\lambda_0}{S} \right)^2\right] ~.
\label{xksSapprox}
\ee
The mean and variance of the above distribution differs from (\ref{meanS}) and (\ref{varS}) by a factor $U_b/s$ and is therefore a good approximation when $U_b \ll U_d$. 


\section{Stationary state distribution when the deleterious mutation rate is smaller than the beneficial one}
\label{app_Udsmall}

For completeness, here we consider the parameter regime in which $U_d < U_b$ within a perturbation theory in $U_d$. We begin by expanding the steady state fraction in a power series in $U_d$ as
\begin{equation}
  x_k=\sum_{n=0}^{\infty} U_d^n \frac{x_k^{(n)}}{n!} ~,
  \label{pert1}
 \end{equation}
 where $x_k^{(n)}$ is the $n${th} derivative of $x_k$ with respect to $U_d$ evaluated at $U_d=0$. When the deleterious mutation rate is zero, as the entire population is in the zeroth fitness class, we have $x_k^{(0)}=\delta_{k,0}$. As a result, the mean ${\cal C}_1^{(0)}= \sum_{k=0}^\infty k x_k^{(0)}=0$. Using  (\ref{pert1}) in (\ref{x0eqn}) and (\ref{xkeqn}) in the steady state and retaining terms to leading order in $U_d$, we obtain
 \bea
 s {\cal C}_1^{(1)} &=& 1- U_b x_1^{(1)} \\
 U_b x_2^{(1)} &=&  (U_b+ {s}) x_1^{(1)} - 1 \\
 U_b x_k^{(1)} &=& (U_b+ {s (k-1)}) x_{k-1}^{(1)} ~,~k \geq 3 ~.
 \eea
 The last equation implies that the steady state fraction is a monotonically increasing function of $k$; however, since each frequency is bounded above by unity and the total fraction must also add up to one, to obtain a sensible result to linear order in $U_d$, the fraction $x_k^{(1)}$ must be zero for all $k \geq 2$. This immediately yields
 \bea
 x_0 &=& 1 -\frac{U_d}{U_b+s} + {\cal O}(U_d^2)\\
 x_1 &=& \frac{U_d}{U_b+s} + {\cal O}(U_d^2) \\
 x_k &=&  {\cal O}(U_d^2) ~,~k \geq 2 ~.
 \eea
 Thus the fraction in the zeroth fitness class decreases linearly with $U_d$ when the deleterious mutation rate is smaller than the beneficial one; in contrast, the fraction $x_0$ decays exponentially or faster with $U_d$ when $U_d > U_b$ (see Sec. ~\ref{sub_largeS} and \ref{sub_smallS}). 
 

\section{Comparison of mutation schemes}
\label{app_reln}

In an infinitely large population of finite diallelic sequences of length $L$ in which the wild type allele mutates with rate $\mu$ and the back mutation occurs with rate $\nu$, the frequency $x_k(t)$ of a sequence with $k$ deleterious mutations and fitness $w_k=-s k$ evolves in continuous time as \cite{Woodcock:1996,John:2015}
\begin{equation}
{\dot x}_k= (k+1) \nu x_{k+1} + (L-k+1)  \mu x_{k-1} - [(L-k) \mu+ k \nu] x_{k} -s (k-\bar k) x_k  ~,
\label{eqbck}
\end{equation}
where $x_{-1}=x_{L+1}=0$. In the limit $\mu, \nu \rightarrow 0$ and $L \rightarrow \infty$, one can define the deleterious and beneficial mutation rate per sequence as $U_d = L \mu$ and $U_b= L \nu$, and rewrite the above equation as 
\begin{equation}
 {\dot x}_k=\epsilon_{k+1}U_b x_{k+1} + (1- \epsilon_{k-1}) U_d x_{k-1} - [(1-\epsilon_{k}) U_d + \epsilon_{k}  U_b] x_{k} - s (k-\bar k)  x_{k} ~,
 \label{eqbck2}
\end{equation}
where $\epsilon_{k}=k/L$. In a well adapted population in which the number of loci carrying the deleterious allele is small, the back mutations to the wild type allele can be ignored. More precisely, when the number of deleterious mutations scales sublinearly with $L$, the fraction $\epsilon_k \to 0$ for an infinitely long sequence and we obtain the model defined by (\ref{x0eqn}) and (\ref{xkeqn}) with $U_b=0$. Similarly, in a maladapted population in which $\epsilon_k \to 1$, the model (\ref{eqbck2}) reduces to the one studied in this article with $U_d=0$. 
The model defined by (\ref{x0eqn}) and (\ref{xkeqn}) thus interpolates between the two limits of the model (\ref{eqbck2}) described above.

\clearpage

\clearpage

\clearpage

\begin{figure}
\includegraphics[width=1 \linewidth,angle=0]{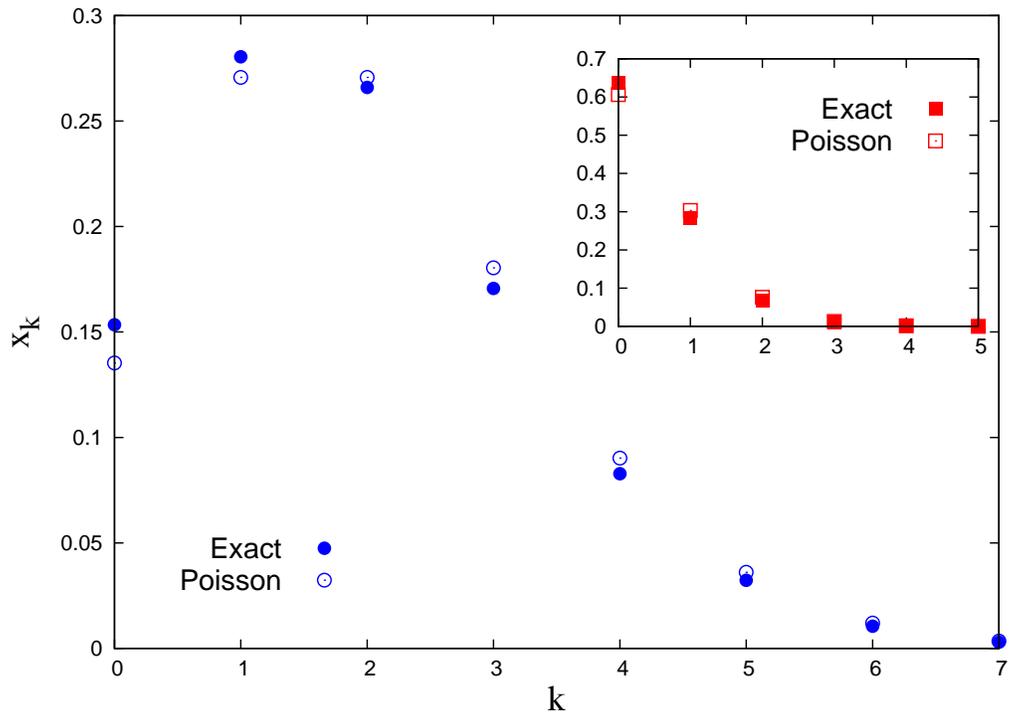}
\caption{Steady state distribution when $ s \gg \sqrt{U_b U_d}$: Comparison of the exact distribution (\ref{xsel_ss}) and Poisson distribution (\ref{Ub0}) for $U_b=0.005, s=0.1, U_d=0.2$ (main) and $U_b=0.01, U_d=0.05, s=0.1$ (inset).}
\label{xklargeS}
\end{figure}

\clearpage

\clearpage

\end{document}